\documentclass[preprint,showpacs,preprintnumbers,amsmath,amssymb]{revtex4}

\usepackage{epsfig}
\usepackage{graphicx}% Include figure files
\usepackage{dcolumn}% Align table columns on decimal point
\usepackage{bm}% bold math

\newcommand{\la}{\langle}
\newcommand{\ra}{\rangle}
\newcommand{\ua}{\uparrow}
\newcommand{\da}{\downarrow}

\newcommand{\om}{\omega}

\begin{document}

%\preprint{APS/123-QED}

\title{Phase diagram and magnetic collective excitations of the 
Hubbard model in graphene sheets and layers}

\author{N. M. R. Peres$^{1,2}$, M. A. N. Ara\'ujo$^{2,3}$ and Daniel Bozi$^{1,2}$}

\affiliation{$^1$Departamento  de F\'{\i}sica, 
Universidade do Minho, P-4710-057, Braga, Portugal,}
\affiliation{$^2$GCEP-Center of Physics, 
Universidade do Minho, P-4710-057, Braga, Portugal,}
\affiliation{$^3$Departamento de F\'{\i}sica, 
Universidade de \'Evora, P-7000, \'Evora, Portugal}

\date{\today}% It is always \today, today,

\begin{abstract}
We discuss the magnetic phases
of the Hubbard model for the   honeycomb lattice both in two
and three spatial dimensions.
A ground state phase diagram is obtained depending on the interaction strength
 $U$ and electronic density $n$.
We find a first order phase transition between ferromagnetic 
regions where the spin is 
maximally polarized (Nagaoka ferromagnetism) and  regions with smaller 
magnetization (weak ferromagnetism). When taking into account the 
possibility of spiral states,
we find that the lowest  critical $U$ is obtained for 
an ordering momentum different from zero. The evolution  of the
ordering momentum with doping is discussed.
The magnetic excitations (spin waves) in the antiferromagnetic insulating phase
are calculated from the random-phase-approximation for the spin
susceptibility. We also compute the spin fluctuation correction to the 
mean field magnetization by virtual emission/absorpion of spin waves. 
In the large $U$ limit, the renormalized magnetization agrees qualitatively
 with the Holstein-Primakoff theory of the Heisenberg antiferromagnet, although
the latter approach produces a larger renormalization.
\end{abstract}

\maketitle

%%%%%%%%%%%%%%%%%%%%
%    SECTION       %
%%%%%%%%%%%%%%%%%%%%%%%%%%%%%%%%%%%%%%%%%%%%%%%%%%%%%%%%%%%%%%%%%%%%%%%%%%%%%%%
\section{Introduction}

The interest in strongly correlated systems in frustrated lattices 
has increased recently because of the possible realization of exotic
magnetic states \cite{anderson}, spin and charge separation in two 
dimensions \cite{matthew}, and the discovery of superconductivity
in Na$_x$CoO$_2$.$y$H$_2$O \cite{tanaka}. Many researchers have 
discussed  superconductivity in non-Bravais lattices, mainly
using self consistent spin fluctuation approaches to the
problem \cite{kuroki,onari,moriya}.
The honeycomb lattice, which is 
made of two inter-penetrating triangular lattices, has received special 
attention 
after  the discovery of superconductivity in MgB$_2$ \cite{nagamatsu2001}.
  Additionally, the honeycomb lattice has been shown to stage many 
different types of
 exotic physical behaviors in magnetism and
the growing experimental evidence of non-Fermi liquid behavior in 
graphite has led to
the study  of electron-electron correlations and quasi-particle
 lifetimes in graphite \cite{gonzalez}.

Around a decade ago,  Sorella and Tossatti \cite{sorella}
found that the Hubbard model in the half-filled honeycomb lattice 
would exhibit a
Mott-Hubbard transition at finite $U$. Their Monte Carlo results were
confirmed by variational approaches and reproduced by other 
authors \cite{martelo,furukawa}. 
As important as the existence of the Mott-Hubbard transition in 
strongly correlated electron systems is the possible realization 
of Nagaoka ferromagnetism. The 
triangular, the honeycomb and the Kagom\'e lattices were studied, but a strong 
tendency for a Nagaoka type
 ground state was found only in non-bipartite lattices (triangular and Kagome) \cite{hanisch}. 
On the other hand, the effect of {\it long range} interactions in half filled sheets of graphite
 was considered from a mean 
field point of view, using an extended Hubbard model. A large region of the phase
 diagram having a charge density wave ground state was found \cite{tchougreeff}. 
More recently, the existence of a new  magnetic excitation in paramagnetic graphite
 has been claimed  \cite{baskaran},  but its existence was reanalyzed by two 
of the present authors \cite{peresI}.

In this work the magnetic phases of the Hubbard model in  the honeycomb lattice are studied.
In addition to the two-dimensional problem we also address  the three-dimensional system
 composed of stacked layers. 
The critical lines associated with instabilities of the paramagnetic phase are obtained
in the $U,n$ plane (interaction versus particle density). Spiral spin phases are also
considered. 
 A  ground state phase diagram containing ferro and antiferromagnetic order is obtained.
Interestingly, we find ferromagnetic regions  with fully polarized spin in the vicinity
of  regions with smaller magnetization. The transitions from one to the 
other are discontinuous. 

We also address the calculation of the 
magnetic excitations (spin waves) in the half-filled antiferromagnetic 
honeycomb layer within the random-phase-approximation (RPA).  
It is known that the Hartree-Fock-RPA theory of the half-filled Hubbard model is 
correct in both weak and strongly interacting
 limits: at strong coupling, the spin wave dispersion  obtained in RPA agrees
 with the Holstein-Primakoff theory for the
Heisenberg model; at intermediate interactions ($U/t\sim 6$), the RPA dispersion shows
 excellent agreement with experiment
\cite{la2cuo4,poznan}. The Hartree-Fock-RPA theory should, therefore, 
be considered as a
 usefull starting point to study the  intermediate coupling regime.  
Starting from  the spin wave spectrum obtained in RPA theory, we  calculate the 
quantum fluctuations correction 
 to the ground state magnetization  arising from  virtual emission/reabsorption 
of spin waves.
 In the strong coupling limit, we find a ground state
magnetization which is about $67\%$ of full polarization. This is not so 
great a reduction as predicted 
by the Holstein-Primakoff theory of the Heisenberg model, which is about $48\%$.

Our paper is organized as follows: in section
\ref{hamilt} we introduce the Hamiltonian and its mean field treatment.
 In section \ref{collective}, 
we discuss the possibility of a well defined magnetic excitation in the paramagnetic phase.
In the ordered phase  at half filling, the spin wave spectrum is computed
and the effect of different hopping terms in the spin wave spectrum is discussed.
In section \ref{instabilities}, the magnetic instability lines
 are obtained and  the possibility of spiral spin phases
for $n<1$ is discussed. The corresponding lowest critical $U$ is determined as 
function of the ordering
wave-vector $\bm q$. 
Section \ref{phasediag} is devoted to  the phase diagram of the system, 
where two different types
of ferromagnetism are found. The first order critical lines separating the 
three ordered phases are
determined.
Section \ref{fluctu} contains a study of the
renormalization of the electron's spectral function  and  magnetization 
by the spin wave  excitations.

%%%%%%%%%%%%%%%%%%%%
%    SECTION       %
%%%%%%%%%%%%%%%%%%%%%%%%%%%%%%%%%%%%%%%%%%%%%%%%%%%%%%%%%%%%%%%%%%%%%%%%%%%%%%%
\section{Model Hamiltonian}
\label{hamilt}

\begin{figure}[ht]
\begin{center}
\epsfxsize=8cm
\epsfbox{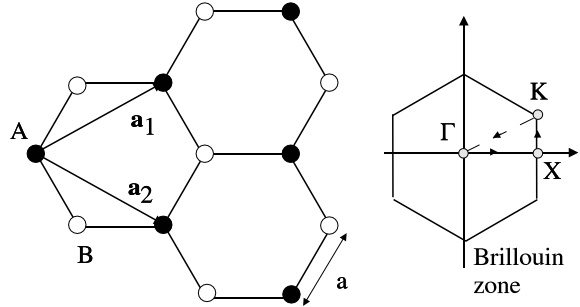}
\end{center}
\caption{Primitive vectors for the honeycomb lattice and 
the corresponding Brillouin zone.}
\label{honey}
\end{figure}

The magnetic properties of the honeycomb lattice is discussed in the context of
the Hubbard model, which  is defined as
\begin{eqnarray}
\hat H=-\sum_{i,j,\sigma}t_{i,j}\hat c^{\dag}_{i,\sigma}\hat c_{j,\sigma}+
U\sum_i
\hat c^{\dag}_{i,\ua}\hat c_{i,\ua}\hat c^{\dag}_{i,\da}\hat c_{i,\da}-
\mu\sum_{i,\sigma}\hat c^{\dag}_{i,\sigma}\hat c_{i,\sigma}\,,
\label{hubbard}
\end{eqnarray}
where  $t_{i,j}$ are hopping integrals, $U$ is the onsite repulsion and $\mu$ 
denotes the chemical potencial.
The honeycomb lattice is not a Bravais lattice since 
there are two atoms per unit cell.
Therefore, it is convenient to define two sublattices, $A$ 
and $B$, as shown in Figure \ref{honey}.

The expressions for the lattice vectors are 
\begin{equation}
{\mathbf{a_1}}= \frac a 2 (3,\sqrt 3,0)\,,
\hspace{1cm}
{\mathbf{a_2}} = \frac a 2 (3,-\sqrt 3,0)\,,
\hspace{1cm}
{\mathbf{a_3}} =  c (0,0,1)\,.
\end{equation}
where $a$ is the length of the hexagon side and $c$ is the interlayer distance.
The reciprocal lattice vectors are given by
\begin{equation}
{\mathbf{b_1}} = \frac {2\pi}{3a} (1,\sqrt 3,0)\,,
\hspace{1cm}
{\mathbf{b_2}} = \frac {2\pi}{3a} (1,-\sqrt 3,0)\,,
\hspace{1cm}
{\mathbf{b_3}} = \frac  {2\pi}{c} (0,0,1)\,.
\end{equation}
The nearest neighbors of an atom belonging to the $A$  sublattice are:
\begin{equation}
{\mathbf{\delta_1}} =  \frac a 2 (1,\sqrt 3,0)
\hspace{0.85cm}
{\mathbf{\delta_2}} = \frac a 2 (1,-\sqrt 3,0)
\hspace{0.85cm}
{\mathbf{\delta_3}} = - a \hat {\bf x}
\hspace{0.85cm}
     {\mathbf{\delta''}} = \pm c \hat {\bf z}
\end{equation}
while the second nearest neighbors (in the plane) 
are: ${\bf \delta'_1}=\pm {\bf a_1}, 
{\bf \delta'_2}=\pm {\bf a_2}, {\bf \delta'_3}=\pm ({\bf a_2}-{\bf a_1})$.
In a broken symmetry state, antiferromagnetic (AF)  order is described by 
the average lattice site  occupation: 
\begin{equation}
<\hat n_{j,\sigma}>=\frac{n}{2} \pm \frac{m}{2}  
\sigma \cos(cQ_z) \left\{\begin{array}{c}
+,j\in A\\ -,j\in B \end{array}\right.
\label{ocupacoes}
\end{equation}
where the $z-$axis ordering vector ${\mathbf{Q}}=(0,0,Q_z)$ 
will be used when studying multi-layers, $n$ denotes the electron 
density, $m$ is the
staggered magnetization, and $\sigma=\pm1$. 
We introduce field operators for each sublattice satisfying the 
usual Fourier transformations:
\begin{equation}
\hat a^\dag_{i\in A,\sigma}=\frac {1}{\sqrt {N}}
\sum_{\bm k}e^{i\bm k\cdot \bm {R_i}}
\hat a^\dag_{\bm k\sigma}\,,\hspace{1cm}
\hat b^\dag_{i\in B,\sigma}=\frac {1}{\sqrt {N}}
\sum_{\bm k}e^{i\bm k\cdot \bm {R_i}}
\hat b^\dag_{\bm k\sigma}\,
\end{equation}
(where  $N$ denotes the number of unit  cells). 
Within a Hartree-Fock decoupling of the  Hubbard 
interaction in (\ref{hubbard}) we 
obtain an effective Hamiltonian matrix 
\begin{equation}
\hat H=\sum_{\bm k\sigma} 
[\hat a^\dag_{\bm k\sigma} \hat b^\dag_{\bm k\sigma}]\left[\begin{array}{cc}
H_{11} & H_{12} \\ H_{21} & H_{22} \\\end{array}\right]
\left[\begin{array}{c} \hat a_{\bm k\sigma} 
\\ \hat b_{\bm k\sigma} \\\end{array}\right],
\label{hamiltHF}
\end{equation}
with matrix elements given by
\begin{equation}
H_{11}=D(\bm k)+U\frac {n-\sigma m} 2 , \hspace{0.5cm} H_{12}=
\phi_{\bm k}=H^\ast_{21} , 
\hspace{0.5cm} H_{22}=D(\bm k)+U\frac {n+\sigma m} 2 
\end{equation}
where
\begin{equation}
\phi_{\bm k}=-t\sum_{\bm\delta}e^{i\bm k\cdot \bm {\delta}} , 
\hspace{0.55cm} D(\bm k)
=\phi'_{\bm k}-2t''\cos(ck_z)-\mu , \hspace{0.55cm} \phi'_{\bm k}=
-t'\sum_{\bm{\delta'}}e^{i\bm k \cdot \bm {\delta'}} \,.
\end{equation}
In the above equations $t$ and $t'$ are the first and second 
neighbor  hopping integrals, respectively, 
while  $t''$ describes interlayer hopping. The  dispersion 
relation for the case where $t'=t''=0$ is
\begin{equation}   
\vert \phi_{\bm k} \vert=t\sqrt{3+2\cos(\sqrt 3 ak_y) + 
4 \cos(3ak_x /2)\cos(\sqrt 3 ak_y /2)}.
\end{equation}
Diagonalization of the effective Hamiltonian yields a  
two band spectrum. The band energies are:
\begin{equation}
E_{\pm}(\bm k)=D(\bm k)+\frac U 2 n \pm 
\sqrt{\Big(\frac {Um} 2\Big)^2 +\vert \phi_{\bm k}\vert^2}.
\end{equation}
Because  there are two sublattices, the Matsubara 
Green's function is a $2\times 2$ 
matrix whose elements are
given by:
\begin{eqnarray}
{\cal G}_\sigma^{aa}(i\omega,{\bm k})&=& \sum_{j=\pm}
\frac{|A_{\sigma,j}|^2}{i\omega - E_j({\bm k})}
\label{gaa}\\
{\cal G}_\sigma^{ab}(i\omega,{\bm k})&=& \sum_{j=\pm}
\frac{A_{\sigma,j}B_{\sigma,j}^*}
{i\omega - E_j({\bm k})}\\
{\cal G}_\sigma^{ba}(i\omega,{\bm k})&=& \sum_{j=\pm}
\frac{A_{\sigma,j}^*B_{\sigma,j}}
{i\omega - E_j({\bm k})}\\
{\cal G}_\sigma^{bb}(i\omega,{\bm k})&=& \sum_{j=\pm}
\frac{|B_{\sigma,j}|^2}{i\omega - E_j({\bm k})}
\label{gbb}
\end{eqnarray}
where the coherence factors are:
\begin{eqnarray}
|A_{\sigma,\pm}({\bf k})|^2 = \frac{1}{2} 
\Big[ 1-\frac{Um\sigma}{2E_\pm({\bf k})}\Big]
&\qquad&
|B_{\sigma,\pm}({\bf k})|^2 = \frac{1}{2} 
\Big[ 1+\frac{Um\sigma}{2E_\pm({\bf k})}\Big]\\
A_{\sigma,\pm}({\bf k})B_{\sigma,\pm}^*({\bf k})
&=&-\frac{\phi({\bf k})}{2E_\pm({\bf k})}
\label{coerenc}
\end{eqnarray}

In the  ferromagnetic (F)  phase, the site occupation 
is the same for both sublattices:   
\begin{equation}
<\hat n_{j,\sigma}>=\frac{n}{2} + \frac{m}{2}  \sigma \qquad
j\in A,B .
\label{ferroocup}
\end{equation}
In this case the quasiparticle energy bands are given by
\begin{equation}
E^{\sigma}_{\pm}(\bm k)=D(\bm k)+\frac U 2 (n-\sigma m)\pm |\phi_{\bm k}|.
\end{equation}

\begin{figure}[ht]
\begin{center}
\includegraphics[angle=0,width=7.5cm]{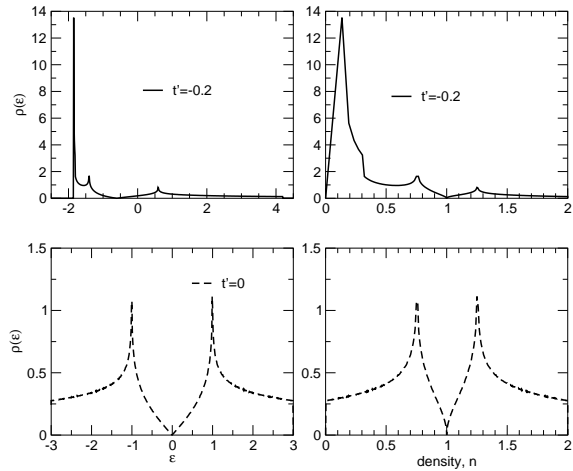}
%\epsfxsize=8cm
%\epsfbox{st_dos.ps}
\end{center}
\caption{
Single particle density of states,
$\rho(\epsilon)$, for independent
electrons in an honeycomb lattice.
The left and right panels show $\rho(\epsilon)$ as function
of energy and electron density, respectively.
The solid line refers to $t'=-0.2$ and the dashed line to $t'=0$.}
\label{dos2d}
\end{figure}

In the paramagnetic phase of the system
the energies and propagators are simply obtained by setting $m=0$ 
in the equations above.
The density of states of  single electrons
is shown in Figure \ref{dos2d} against particle density and energy. 
In the two upper panels we have included a second-neighbor 
hopping while in the two 
lower panels only  nearest neighbor coupling is considered.
An  important feature is that $\rho(\epsilon)$ vanishes 
linearly with $\epsilon$ as we 
approach the half filled limit, both for $t'=0$ and $t'\ne 0$. 
This is related to the $K$-points
of the Brillouin Zone (see Figure \ref{honey}), where the 
electron dispersion becomes linear: 
$$
E({\bf k})  \approx \pm t \frac{3a}{2}|d{\bf k}|
$$  
($d{\bf k}$ denotes the deviation from the $K$-point). 
This dispersion is called the ``Dirac cone''.

%%%%%%%%%%%%%%%%%%%%%%%%%%%%%%
%   COLLECTIVE EXCITATIONS   %
%%%%%%%%%%%%%%%%%%%%%%%%%%%%%%%%%%%%%%%%%%%%%%%%%%%%%%%%%%%%%%%%%%%%%%%%%%%%%%%
\section{Collective excitations at half filling}
\label{collective}

The magnetic  excitations are obtained from  the poles of the 
transverse spin susceptibility tensor, $\chi$,
which is definded, in Matsubara form, as
\begin{equation}
\chi^{i,j}_{+-}(\bm q,i\omega_{n})=
\int_0^{1/T} d\tau e^{i\omega_{n}\tau}\la T_{\tau} 
\hat S^{+}_i(\bm q,\tau)\hat S^{-}_j(-\bm q,0) \ra
\end{equation}
where $i,j=a,b$ label the two sublattices (not lattice points)
and $S^{+}_i(\bm q),\  S^{-}_j(\bm q)$ denote the
spin-raising and lowering operators for each sublattice.

In  the paramagnetic, F, or AF phases, the  zero order susceptibility is 
just a simple bubble diagram with the Green's functions given
in equations  (\ref{gaa})-(\ref{gbb}): 
\begin{equation}
\chi_{+-}^{(0)i,j}(\bm q,i\omega_n)=
-\frac T N \sum_{\bm k ,\omega_m} {\cal G}_{\ua}^{ji}(\bm k,i\omega_{n}) 
{\cal G}_{\da}^{ij}(\bm {k-q},i\omega_n - i\omega_m)
\label{chi0}
\end{equation}
Going beyond mean-field, the random-phase-approximation 
(RPA) result   for the susceptibility tensor is obtained
 from the Dyson equation 
\begin{equation}
\chi = \chi^0 +  U  \chi^0 \chi\ 
\Rightarrow \ \chi = \Big[ \hat I -  U\chi^0 \Big]^{-1} \chi^0
\end{equation}
where $\hat I$ denotes the $2\times 2$ identity matrix. 
The poles of the susceptibility tensor, corresponding to the
magnetic excitations, are then obtained from the condition:
\begin{equation}
{\rm Det} \Big[ \hat I -  U\chi^0 \Big] = 0.
\label{det}
\end{equation}
We note that the tensorial nature of the spin susceptibility 
is a consequence of there being two sites
per unit cell and is not related to the magnetic order in the system.

%%%%%%%%%%%%%%%%%%%%%%%%%%%%%%
%   THE PARAMAGNETIC PHASE   %
%%%%%%%%%%%%%%%%%%%%%%%%%%%%%%%%%%%%%%%%%%%%%%%%%%%%%%%%%%%%%%%
\subsection{Magnetic excitations in a single paramagnetic layer}

Here we discuss the possibility of existence of  magnetic excitations 
 in a single honeycomb paramagnetic layer. 
Our interest in this problem stems from a recent claim, by 
Baskaran and  Jafari \cite{baskaran},
who  recently proposed the existence of a neutral spin 
collective mode in graphene sheets. In the 
calculations of Ref. \cite{baskaran}  a half-filled Hubbard model 
in the honeycomb lattice (with  $t'=t''=0$) was considered but
 the tensorial character  of the susceptibility was neglected
\cite{peresI}.  
Since inelastic neutron scattering can be used to study this 
spin collective mode in graphite,
we decided to re-examine this problem taking into account 
the tensorial nature of the transverse spin susceptibility.

Collective magnetic modes with frequency $\omega$ and 
momentum ${\bm q}$ are determined from the condition (\ref{det}) 
after performing the analytic continuation $i\omega 
\rightarrow \omega + i0^+$. The determinant is given by
\begin{equation}
D_{+-}(\bm q,\omega)=1-2U \chi_{+-}^{(0)aa}+U^2 
\Big[ (\chi_{+-}^{(0)aa})^2-\chi_{+-}^{(0)ab}\chi_{+-}^{(0)ba}\Big],
\label{pole}
\end{equation}
where we have taken into account that in a  paramagnetic 
system  $\chi_{+-}^{(0)aa}=\chi_{+-}^{(0)bb}$.
Below the particle-hole continuum of excitations, the 
spectral (delta-function contributions) part in
$\chi^{(0)ij}_{+-}(\bm q,\omega  +i0^+)$ vanishes and 
there is the additional relation 
$\chi_{+-}^{(0)ba}=(\chi_{+-}^{(0)ab})^\ast$.
Collective modes are only well defined outside the particle-hole 
continuum (inside the continuum they become Landau damped). 
 We searched\cite{peresI} for well defined 
magnetic modes, $\omega(\bm q)$,  below the continuum
of particle-hole excitations, and found no solutions for any 
value of the interaction $U$. 
In  Figure 1 of Ref. \cite{peresI} we plot $D_{+-}(\bm q,\omega)$ for eight different 
$\bm q$-vectors and $\omega$ ranging from zero to the point where the
particle-hole continuum begins.  Our analysis reveals that the full 
tensorial structure 
of the Hubbard model's RPA susceptibility in the honeycomb
lattice does not predict  a collective magnetic mode. 

%%%%%%%%%%%%%%%%
% Sub section  %
%%%%%%%%%%%%%%%%
\subsection{Spin waves in the antiferromagnetic layer}

\begin{figure}[ht]
\begin{center}
\epsfxsize=8cm
\epsfbox{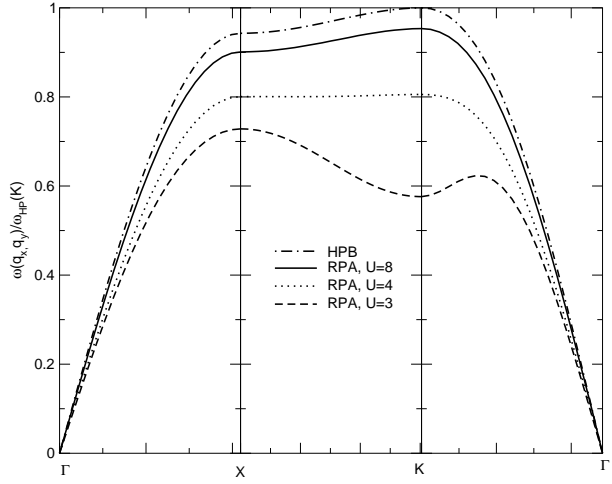}
\end{center}
\caption{Spin-wave excitation spectrum for several values of $U$. 
The dashed-dotted line gives the Holstein-Primakoff result 
for the Heisenberg antiferromagnet in the honeycomb lattice. }
\label{afsw}
\end{figure}

\begin{figure}[ht]
\begin{center}
\epsfxsize=8cm
\epsfbox{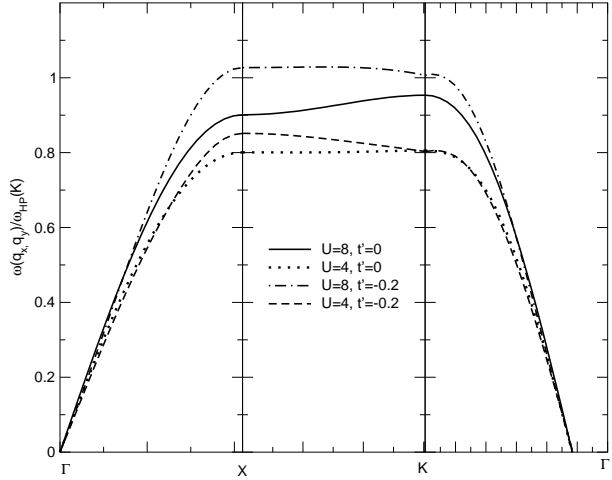}
\end{center}
\caption{Spin-wave excitation spectrum for several values 
of $U$ and $t'\ne 0$.}
\label{afswtp}
\end{figure}

The  spin wave dispersion  $\omega(\bm q)$ for the AF layer with 
one electron per site can be
obtained from equations (\ref{chi0}) and (\ref{det}) using  
expressions (\ref{gaa})-(\ref{gbb}) 
 for the propagators. 
Spin wave spectra, for different values of second-neighbor 
hopping,  $t'$, are ploted in Figures \ref{afsw} and \ref{afswtp}. 
In the large $U$ limit, spin wave energies  agree
with those obtained from the Holstein-Primakoff theory of 
the Heisenberg model. 
We give an analytical  derivation of this limit 
in Appendix \ref{appendsuslarge}.
The Holstein-Primakoff result for  the Heisenberg model in 
the honeycomb lattice, which  is 
derived in Appendix \ref{appendHP}, can be written as
\begin{equation}
\omega_{HP}(\bm q)= JS\sqrt{z^2-\vert\phi(\bm q)\vert^2}\,.
\end{equation}
This result can be mapped on the  Hubbard model provided that 
$J=4t^2/U$ and $S=1/2$. 
Figure \ref{afsw} shows the spin wave energies for the 2D 
lattice ($t'=0$) along a closed path in the Brillouin Zone. 
Energies in Figure \ref{afsw} are normalized by  the 
Holstein-Primakov result at the $K$-point, $\omega_{HP}(K)$ 
(see Figure \ref{honey}). 
It can be seen that  the results  for $U=8$ are very close 
to the asymptotic behavior of the RPA,
 whereas, for smaller $U$, the spin wave energy is reduced. 
 The effect of $t'$ on $\omega(\bm q)$ is depicted in Figure \ref{afswtp}. 
It is of particular interest the
fact that the dispersion along the $X-K$ direction is almost 
absent for $U\ge 4$. The
presence of $t'$ does not change this effect.

%%%%%%%%%%%%%%%%%%%%%%%%%%%%%%%%%%%
%     MAGNETIC INSTABILITIES      %
%%%%%%%%%%%%%%%%%%%%%%%%%%%%%%%%%%%%%%%%%%%%%%%%%%%%%%%%%%%%%%%%%%%%%%%%%%%%%%%
\section{Magnetic instabilities}
\label{instabilities}
\begin{figure}[ht]
\begin{center}
\epsfxsize=10cm
\epsfbox{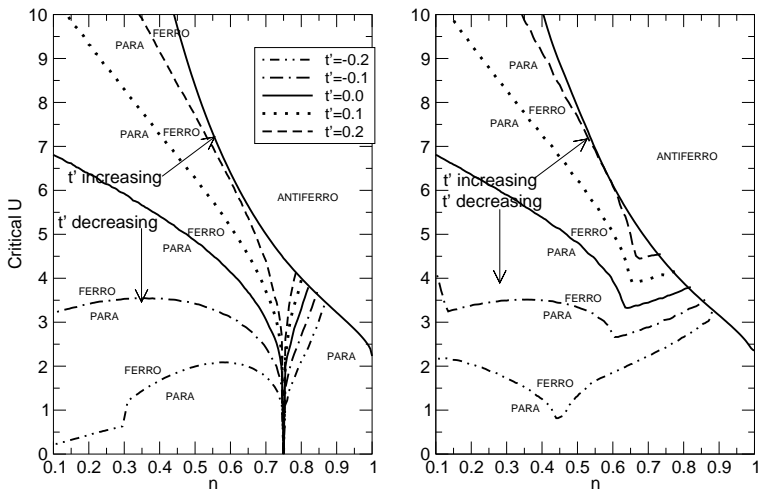}
\end{center}
\caption{
{\bf left panel:} Effect of $t'$ on the instability lines, as determined from the 
 equation (\ref{Ucrit}), for a single honeycomb layer.
{\bf right panel:} Effect of $t'$ on the instability lines, as determined from 
equation  (\ref{Ucrit}), for a layered honeycomb.
This panel differs from the other inasmuch a small $t''=0.1$ hoping term
was included coupling the 2D layers.}
\label{2Dcrit}
\end{figure}

The magnetic instabilities in the paramagnetic phase can be obtained from the 
divergence of  the RPA susceptibilities
at critical values of the interaction, $U_c$, driving the 
system towards a magnetically ordered phase. 
At  a given electron density $n$
 we always find two instability solutions, one  ferromagnetic  and one 
antiferromagnetic.
One of these solutions minimizes the  free energy. 
Since $U_c$ is determined from 
 $D_{+-}(\bm q,0)=0$,  taking into account 
that $\chi_{+-}^{(0)aa}=\chi_{+-}^{(0)bb}$ and
 $\chi_{+-}^{(0)ab}=(\chi_{+-}^{(0)ba})^{\ast}$ in 
the paramagnetic phase, we obtain:
\begin{equation}
U_c = \frac 1 {\chi_{+-}^{(0)aa}\pm \vert \chi_{+-}^{(0)ab} \vert}.
\label{Ucrit}
\end{equation}
Figure   \ref{2Dcrit} shows $U_c$ obtained from   
the  static uniform susceptibilities
 ($\bm q =\bm 0$ and $\omega=0$), as a function of  
electron density for various  values of $t'$. 
Detailed equations for the  instability lines are given in 
Appendix \ref{appendsusuc}. 
The left panel of Figure \ref{2Dcrit} refers
 to the 2D case, corresponding to a single honeycomb layer, 
whereas the right panel  refers to the 3D system 
with a constant interlayer hopping $t''=0.1$. The Van-Hove 
singularity (associated with  the $X$ point) 
plays an important 
role  at  density $n=0.75$  in the 2D case, 
independently of $t'$.
%\vspace{0.7cm}

As we have already mentioned, the two solutions of Eq. (\ref{Ucrit}) 
correspond to two different 
magnetic transitions, one between a paramagnetic phase and a ferromagnetic 
phase and 
another between a paramagnetic phase and 
an antiferromagnetic phase. That this is so can easily be confirmed by 
solving the self-consistent equations for the ferromagnetic and the 
antiferromagnetic magnetizations, respectively, derived from the
HF Hamiltonian (\ref{hamiltHF}). By minimizing the free 
energy with respect to magnetization,  
one finds the following expressions for
ferro and antiferromagnetic magnetizations
\begin{equation}
m_F=\frac 1 {2N} \sum_{\bm k \sigma} \sigma (f(E^{\sigma}_+)
+f(E^{\sigma}_-))\,, 
\hspace{1.0cm} m_{AF} = \frac 1 N \sum_{\bm k} 
\frac {\vert \zeta_{\bm k} \vert} 
{\sqrt{1+\zeta^{2}_{\bm k}}} (f(E_-)-f(E_+))\,,
\label{selfconsmagn}
\end{equation}
where $f(x)$ is the Fermi function and 
$\zeta_{\bm k}=Um_{AF}/(2 \vert \phi_{\bm k} \vert)$.
Letting both $m_F$ and $m_{AF}$ approach zero, one obtains 
the same lines as those in Figure \ref{2Dcrit}. 
Generally speaking, for electron densities 
lower than $0.85$, the value of $U_c$ that saparates the 
paramagnetic region from the ferromagnetic 
region is lower than the corresponding value of $U_c$ 
separating the paramagnetic region from the 
antiferromagnetic region.
The critical $U$ associated with  the ferromagnetic instability
increases with  $t'$.
The size of the paramagnetic region in Figure \ref{2Dcrit} 
increases with $t'$. On the other hand, for $t'=0.2$, we see 
that the critical line for the  
ferromagnetic region is very close the 
critical line 
of the antiferromagnetic region. Therefore, the 
ferromagnetic region is progressively shrinking  with increasing $t'$.
 If we now  turn to densities larger than $0.85$, we find 
that  the antiferromagnetic critical line is the one with lowest $U_c$. 
However, in contrast to lower densities, the 
antiferromagnetic critical line  hardly changes when varying $t'$.  
This description applies equally well to the single honeycomb 
layer and weakly coupled layers, even though the quantitative functional 
dependence of $U_c$ on  $n$ is different in the two cases, the main 
difference coming from the van Hove singulary present in the 2D case.
At finite temperature the van Hove singularity is rounded off
and the 2D phase diagram will be much more similar
to the 3D case. We therefore, consider that
a weak 3D inter-layer coupling does not qualitatively modify
the conclusions valid for the 2D case.

Besides collinear spin phases, the system may also present non-collinear -- spiral --
spin phases in some regions of the phase diagram. We now study what are the changes
in the critical $U$ values determining the instability of the
paramagnetic phase if we allow for non-collinear ground states,
since it is well known that the 
Hubbard model on bipartite and 
non-bipartite lattices can have the lowest  $U_c$ for spiral 
spin phases \cite{hanisch,krishnamurthy,kampf} for some electronic densities. 
In a spiral state, the   spin expectation value 
 at site $i$, belonging to sublattice $\nu=a,b$, is given by
\cite{subir}  
\begin{equation}
\la \bm {S^{\nu}_i} \ra = \frac {m_{\nu}} 2 (\cos(\bm q \cdot \bm {R^{\nu}_i}),
\sin(\bm q \cdot \bm {R^{\nu}_i})).
\end{equation}
If $\bm q \ne \bm 0$, the ferromagnetic and antiferromagnetic spin 
configurations become twisted. We shall refer to 
the twisted $\bm q \ne \bm 0$ 
configurations as '$F_q$' whenever $m_A=m_B$, and '$AF_q$' whenever $m_A=-m_B$. 
The criterion for choosing the $\bm q$-vectors is taken directly from the 
geometry of the lattice by requesting a constant angle
 between spins 
on  neighboring sites, i.e. $\bm q \cdot \bm {\delta_{1}}=\bm q 
\cdot \bm {\delta_{2}}=-\bm q \cdot \bm {\delta_{3}}$.
\begin{figure}[ht]
\begin{center}
\epsfxsize=8cm
\epsfbox{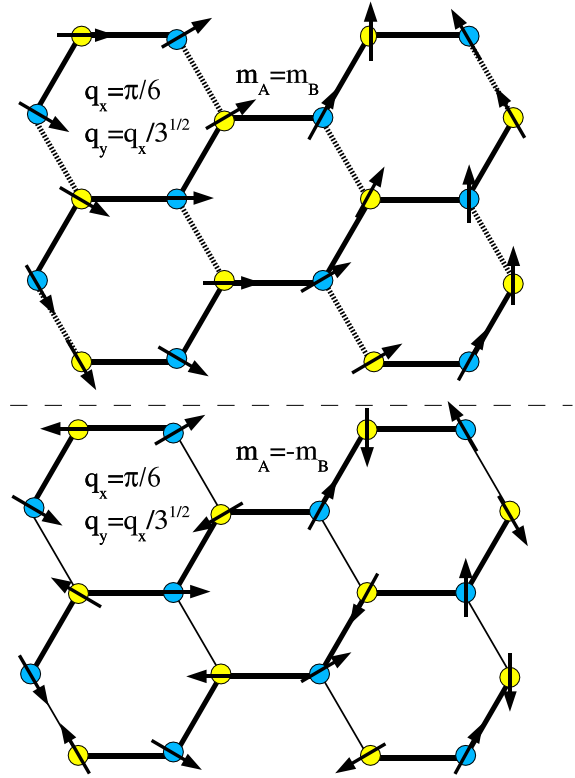}
\end{center}
\caption{(color on line)
$F_q$ (upper) and $AF_q$ (lower) spin configurations for 
$q_x=\frac {\pi} 6$.}
\label{example_1_q=pi/6}
\end{figure} 
 \begin{figure}[ht]
\begin{center}
\epsfxsize=8cm
\epsfbox{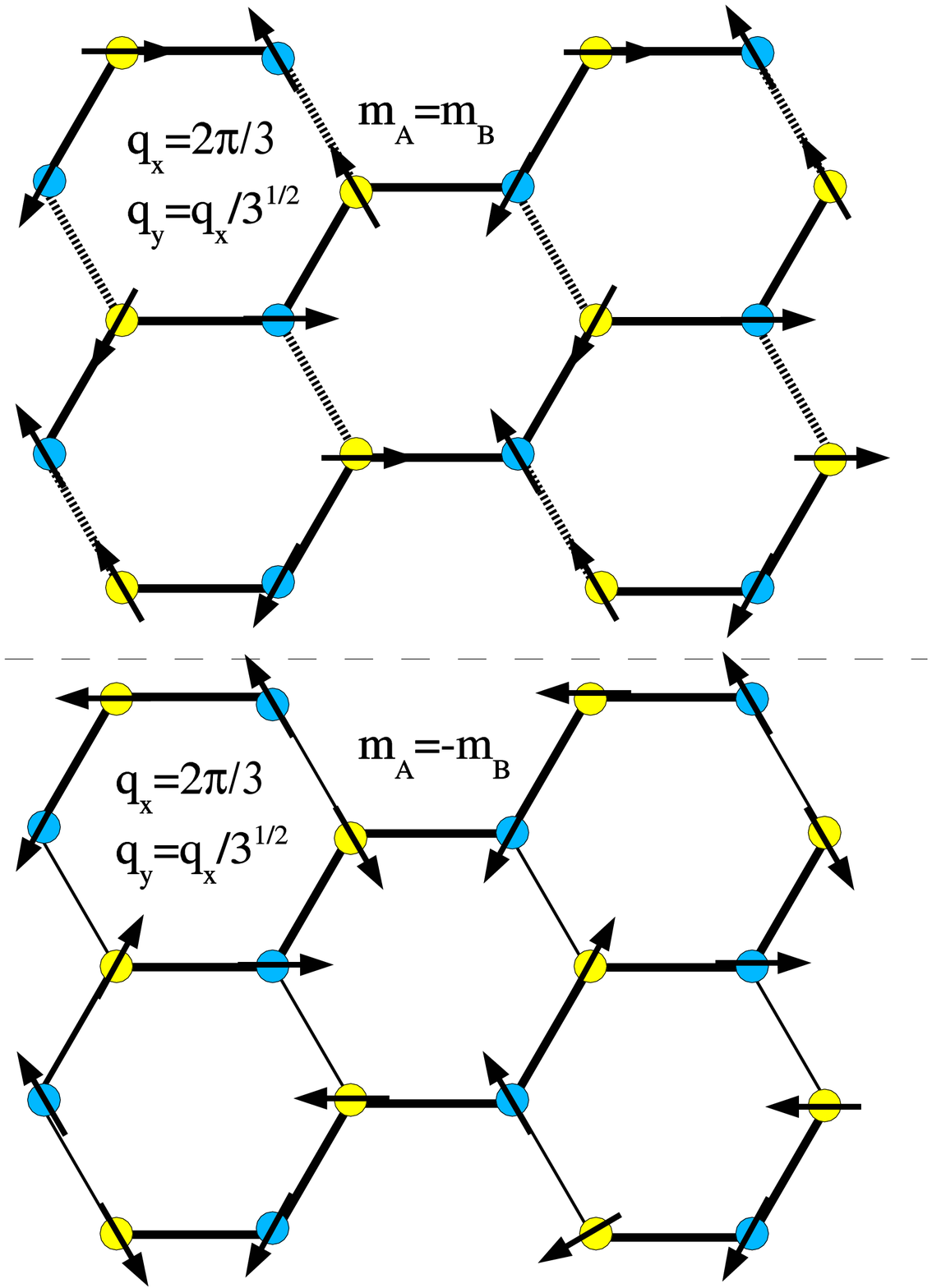}
\end{center}
\caption{(color online)
$F_q$ (upper) and $AF_q$ (lower) spin configurations for 
$q_x=\frac {2\pi} 3$.}
\label{example_2_q=2pi/3}
\end{figure} 
Unfortunately, however, this cannot be achieved in the honeycomb lattice 
with only one $\bm q$-vector. The closest one can get to a 'true' spiraling 
state is by letting $\bm q \cdot \bm {\delta_{1}}
=-\bm q \cdot \bm {\delta_{3}}$ 
(or equivalently,  $\bm q \cdot \bm {\delta_{2}}
=-\bm q \cdot \bm {\delta_{3}}$), 
which implies that $\bm q=(q_x,q_y)
=q_x(1,\frac 1 {\sqrt{3}})$ ($\bm q
=q_x(1,\frac {-1} {\sqrt{3}})$). For the moment we let $q_z$ 
be zero which means that we consider identical layers. The condition $\bm q \cdot 
\bm {\delta_{1}}=-\bm q \cdot \bm {\delta_{3}}$ 
means that the increase in spin angle between two 
lattice sites in the $-\bm {\delta_{3}}$ 
direction is the same as the increase in spin 
angle between two lattice sites in the $\bm {\delta_{1}}$ 
direction. 
There is no increase in the spin angle in the $\bm {\delta_{2}}$ direction. 
Examples of the 
spin-configurations obtained in this way are 
shown in Figures \ref{example_1_q=pi/6} and \ref{example_2_q=2pi/3}. 
Several notes are in order at this stage. First, although we do not have
a 'true' spiraling state over the whole lattice, 
we do have a spiraling configuration in the $-\bm {\delta_{3}}$ 
and $\bm {\delta_{1}}$ directions, as
can be seen  from the Figures \ref{example_1_q=pi/6} 
and \ref{example_2_q=2pi/3}, going from  the lower left to upper  right.
Secondly, when travelling along
 the $\bm {\delta_{2}}$ direction, the spin angles do 
not increase. Instead,  neighboring  spins 
in this direction are always aligned ferromagnetically when $m_A=m_B$, 
and antiferromagnetically when $m_A=-m_B$. 
However, two successive $\bm {\delta_{2}}$ bonds ('sliding down' the lattice 
from  left to right) 
have the same increase in spin angle as any two neighbors connected by 
$-\bm {\delta_{3}}$ or $\bm {\delta_{1}}$.
The $\bm q$-vector (i.e. the spin configuration) that a 
system with a given density 
would prefer is the one with  the lowest value of $U_c(\bm q)$. 
In Figure \ref{minangle}
 we present a curve showing  the $\bm q$ vectors 
that minimize $U_c (\bm q)$, as  functions of 
particle density 
$n$. We consider discrete values $q_x=i \frac {\pi} {12}$ with  $i=0,1,...,12$. 
The dependence on $t'$ 
is overall the same as that discussed for $\bm q = \bm 0$ 
(for example, the shrinking effect with increasing $t'$ is also seen here). 
There is no reason to restrict  $\bm q$ to integer 
multiples of $\frac {\pi} {12}$, 
other than a pure computational one. By performing the same 
calculation with more $\bm q$-vectors, 
the 'step function' like appearance of the lower graphs of Figure \ref{minangle} 
can be smoothed out. 
Our analysis is sufficient, however, to get an insight into 
how the $\bm q$ vectors (which minimize $U_c$)
vary with $n$.

\begin{figure}[ht]
\begin{center}
\epsfxsize=8cm
\epsfbox{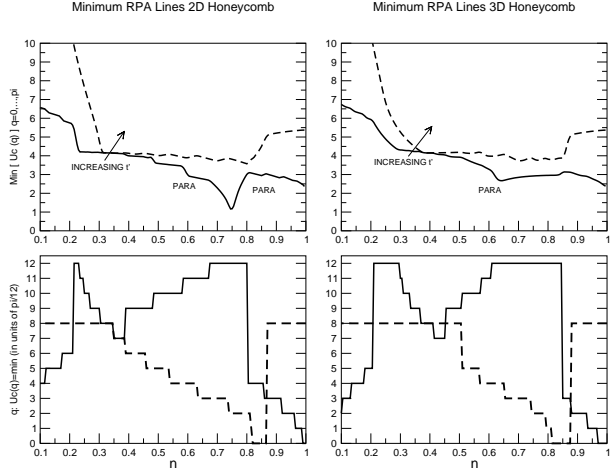}
\end{center}
\caption{ 
The
upper panels show  the minimum  $U_c (\bm q)$ according to 
Eq. (\ref{Ucrit}). The solid line separates
the  paramagnetic phase from 
magnetically ordered phases, while the dashed line  
separates differently ordered magnetic phases.
The lower panels show the  $q_x$ component of the ordering vector
(corresponding to the minimum  $U_c$) as 
function of electron density $n$.}
\label{minangle}
\end{figure} 
The solid line limiting the paramagnetic region is shown in 
the lower graphs of Figure \ref{minangle}). 
We see that the behavior of $q_x$,  as  function of $n$, 
is almost the same for the 2D and 3D cases. As the system 
approaches half filling, the prefered 
spin configuration approaches that with $\bm q= \bm 0$.
In a doped system, however,  minimization of $U_c(\bm q)$
is attained for a  non-zero $\bm q$. 
It is also seen that  the dependence of $\bm q$ on $n$ is not monotonic. 
Either in  2D or  3D,  $q_x$ goes all the way  from $0$ (at $n=1$) 
to $\pi$, displaying  two local maxima (and a local minimum 
in between) as $n$ ranges  from $1$ towards $0$. 

The value of $q_x$ reaches  a 
local minimum at $q_x=\frac {7\pi} {12}$, 
at $n=0.37$ (in 2D) or at $n=0.45$ (in 3D). For even lower densities, $q_x$ 
attains another maximum at $q_x=\pi$, which 
means that the spins of any two nearest neighbors, 
in the $-\bm {\delta_{3}}$ and $\bm {\delta_{1}}$ 
directions, point exactly in opposite directions to 
each other. 

The same type of behavior is seen  also for 
the critical line separating  magnetically 
ordered phases (dashed line). Again,  
the   2D and the 3D 
cases are very similar to each other. For densities 
around $0.30-0.35$ (2D) and $0.35-0.40$ (3D), we have 
$q_x=\frac {8\pi} {12}$ 
yielding the lowest $U_c$. Moreover, the solid and the dashed 
lines coincide, illustrating the previously 
mentioned ferromagnetic 'shrinking out' effect. 
In other words, for $q_x=\frac {8\pi} {12}$, the 
two solutions of $U_c(\bm q)$ almost coincide for all $n$, 
leaving only a thin strip of ferromagnetism 
between the paramagnetic and the antiferromagnetic regions. 
Although this is true for all $n$, 
it is only for $n=0.37-0.40$ (2D case) and $n=0.35-0.37$ (3D case) 
that $U_c(q_x=\frac {8\pi} {12})$ is minimum.

So far, our analysis has been restricted to $\bm q$-vectors 
lying in the $x-y$ spin plane. This means 
that two inter-layer neighbors have the same spin. If we 
now consider  neighboring layers  with  opposite spin, we put $q_z=\pi$. 
At half filling, the lowest $U_c(0,0,\pi)=2.04$ 
limiting the paramagnetic region is lower than 
the corresponding  $U_c(0,0,0)=2.35$, 
independently of $t'$. 
Moreover, for $n=1$, $U_c(q_x,\frac {q_x} {\sqrt{3}},\pi)$ is always 
lower than $U_c(q_x,\frac {q_x} {\sqrt{3}},0)$
for any $q_x$, showing that, at half filling, we should 
expect  antiferromagnetic ordering along the $z$-direction.

The study above was focused on the second order instability lines, both in the
case of collinear and spiral spin phases, being clear
 that spiral states have a lower critical $U-$ value, over a large
range electronic densities. It is instructive to compare
our results with those of Ref. \cite{hanisch}.
Looking at  Fig. 2 of Ref.\cite{hanisch}  we see that for the
triangular lattice there are some finite regions where the more stable ground states
correspond to spiral states. These regions are located
at electronic densities smaller than 0.5 and larger than 0.8. 
Since the honeycomb lattice consists of
two inter-penetrating triangular lattices
we expect the same type behavior, at least
at the qualitative level. That is, we do expect to have finite
regions of the phase diagram where  spiral phases have the lowest energy.
Also, in Ref. \cite{hanisch} the authors do not discuss the full phase diagram
of the Hubbard model in the honeycomb lattice,
as we do in next section. They are primarily
interested in the stability of the Nagaoka state.
Their study is done using three different approaches
(i) The Hartree single flip ansatz; (ii) the SKA Gutwiller ansatz;
(iii) the Basile-Elser ansatz. A comparison can be established between
the  the Hartree single flip ansatz which roughly speaking,
produces a straight line for all densities at the on-site
Coulomb interaction 
$U\sim 5$, and  our self consistent Hartree-Fock study.
If we forget, for a moment, the van Hove singularity, both results
are qualitatively the same for $n$ up to 0.8. Above this value
our Hartree-Fock analysis, forgetting about the existence of the 
antiferromagnetic phase, predicts a very strong
increase of the critical $U$ value (not shown
in Fig. \ref{2Dcrit}, since the AF phase
presents the lowest critical $U$-value), in agreement with
the SKA ansatz. This behavior is not captured by the
 the Hartree single flip ansatz. It seems that our study
interpolates between the  Hartree single flip ansatz
for low densities and the SKA ansatz for densities above 0.8.
Quantitatively there are differences between the two studies,
which are understandable on the basis of the different
types of proposed ground states.

%%%%%%%%%%%%%%%%%
% PHASE DIAGRAM %
%%%%%%%%%%%%%%%%%%%%%%%%%%%%%%%%%%%%%%%%%%%%%%%%%%%%%%%%%%%%%%%%%%%%%%%%%%%%%%%
\section{Phase diagram}
\label{phasediag}

\begin{figure}[ht]
\begin{center}
\epsfxsize=8cm
\epsfbox{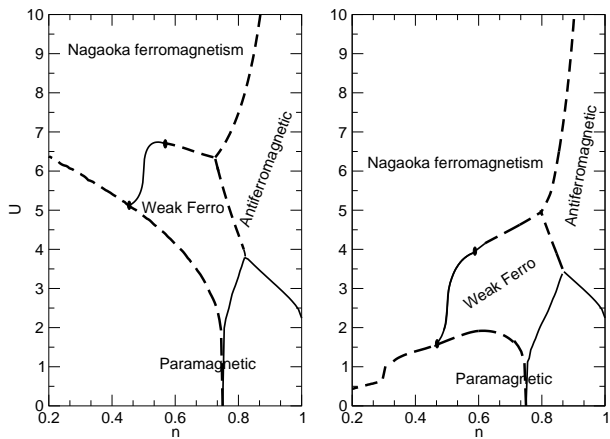}
\end{center}
\caption{{\bf left panel:}
Ground state phase diagram of the Hubbard model in the  $(n,U)$ plane
for a single layer with $t'=0$.
{\bf right panel:} Ground state phase diagram of the Hubbard 
model in the  $(n,U)$ plane
for a single layer with $t'=-0.2$.
In both cases  dashed and continuous lines represent first  
and second order transitions, respectively.}
\label{phase}
\end{figure}

As we mentioned in the previous section, the study of Ref. \cite{hanisch}
is mainly concerned with the stability of the Nagaoka state,
and in the previous section
we studied the  values of the  
Hubbard interaction associated with 
instabilities of the paramagnetic system.
 The transition from the paramagnetic to a magnetically 
ordered state is  determined by the lowest $U_c$.
Since we have found the possibility of having, at least, two 
(ferro and antiferro)
different types of ground states, then 
in the case where interaction is stronger than both critical  values, 
we need to address the problem of  
 competition between the two ordered phases. 
The phase with the lowest free energy is the one prefered by the system. 
In this section we
 restrict ourselves to the study of a single layer but we 
shall consider different
band structures. Spiral states will not be considered, since
we are most interested in a weak ferromagnetic phase showing up
in region of the phase diagram where the studies of Ref. \cite{hanisch}
suggest that the collinear ferromagnetic (fully polarized) phase 
should be the most stable one. 
In the ferromagnetic phase we distinguished two types of ferromagnetic 
ground states:
the Nagaoka ground state, with a maximally polarized spin
 ($m_F=n$), and a weak ferromagnetic state with $m_F<n$. 
The order parameter and 
free energies were obtained from the mean
 field Hamiltonian (\ref{hamiltHF}). Figure \ref{phase} shows the 
ground-state  $(n,U)$-phase-diagram of the model. 
%\vspace{0.5cm}

The effect of  $t'$ on the phase diagram can be 
seen in right panel of Fig.  \ref{phase}.
In Figure \ref{phase}  the dashed lines 
represent first-order phase transitions, where
the order parameter do not vanish smoothly,
while continuous lines represent second order transitions,
where the order parameter vanishes smoothly, but its
first derivative is discontinuous. 
In both cases ($t'=0$ and $t'\ne 0$) we find a finite region of 
weak ferromagnetism.
 In general the Nagaoka phase  
is more stable for large $U$. 
The weak ferromagnetic phase is separated from the Nagaoka phase
by  first or second order transition lines, depending
on the path followed on $(U,n)$ diagram. The second order transition
manifests itself through a discontinuity of the derivative
of the magnetization with respect to $U$.  
 At $n=0.75$ the instability line towards the ferromagnetic phase shows a dip 
(pronouced if $t'=0$), 
which is due to  the logarithmic van-Hove singularity at $n=0.75$. 
A negative $t'$ produces two 
effects on the phase diagram:
 (i) the instability line towards the $F$ phase moves downwards; 
(ii) the point where the  instability lines towards  $F$ and $AF$ 
meet moves to larger $n$.
 Similarly to what was found in the previous section, 
the overall effect of $t'$ is to modify the ferromagnetic 
region of the phase diagram.
 Further, for negative $t'$ we expect collinear
ferromagnetism to exist over a large phase of the phase diagram
relatively to the case $t'\ge 0$, since it is well known that
a negative $t'$ stabilizes the ferromagnetic phase.
On the other hand we don't expect the phase diagram
presented in this section  to be fully accurate
for low densities, where the findings of Ref. 12 should apply.

 The first order critical lines do separate
two different ferromagnetic (or ferromagnetic from antiferromagnetic) regions,
in what concerns the total magnetization. 
In view of the results  published
in  Ref. \cite{burgy},
where a first order transition between the two competing
phases is transformed by disorder
into two second order phase transitions,  we expect the
same behavior to apply here, that is, disorder may
change the order of the transition, since the arguments 
put forward in Ref. \cite{burgy} are of very general
nature. It would be very interesting to study
whether the introduction of disorder in the system
 could change the nature of the first order transitions. 

%%%%%%%%%%%%%%%%%%%%%%%%%%%%
%   QUANTUM FLUCTUATIONS   %
%%%%%%%%%%%%%%%%%%%%%%%%%%%%%%%%%%%%%%%%%%%%%%%%%%%%%%%%%%%%%%%%%%%%%%%%%%%%%%%
\section{Quantum fluctuations}
\label{fluctu}

This section is devoted to  the calculation of quantum fluctuation corrections 
to the magnetization.
An analogous calculation for the Hubbard model in the square 
lattice in the  $t/U\rightarrow 0$ limit 
was skecthed by Singh and Te\v{s}anovi\'c.\cite{singh}

\begin{figure}[ht]
\begin{center}
\epsfxsize=8cm
\epsfbox{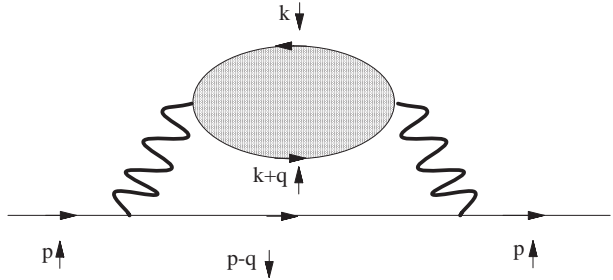}
\end{center}
\caption{The self-energy for a $\uparrow$-spin electron.
 The bubble represents  the transverse susceptibility
computed in RPA.}
\label{trovao}
\end{figure}

The computation of the renormalized staggered magnetization requires the 
evaluation of the Feynman diagram shown in Figure (\ref{trovao}), 
which shows the second order
(in the interaction $U$) contribution to the self-energy.    
The diagram describes the emission and later absorption of a 
spin wave by an up-spin electron. 
The emission and absorption processes are accompanied by 
electron spin reversal.  
This effect, consisting of virtual spin flips, is  going to 
renormalize the staggered magnetization.
The spin-$\ua$ electron Green's function is 
$$
{\cal G}_\ua({\bf p} , i\omega) = {\cal G}_\ua^0({\bf p} , i\omega) 
+ {\cal G}^0_\ua({\bf p} , i\omega)
 \Sigma_\ua({\bf p} , i\omega) {\cal G}_\ua({\bf p} , i\omega)\,,
$$
hence,  ${\cal G}^{-1}=[{\cal G}^{(0)}]^{-1} - \Sigma^{-1}$. 
Here, ${\cal G}^0$ 
denotes the Hartree-Fock Green's functions
matrix appearing in equations (\ref{gaa})-(\ref{gbb}). 
The self-energy matrix is given by
\begin{equation}
\Sigma_{\ua}^{ij}({\bf p} , i\omega) = U^2\frac{T}{N}\sum_{i\Omega,{\bf q}}
{\cal G}_\da^{(0)ij}({\bf p - q} , i\omega - i\Omega)
\chi_{-+}^{(RPA)ij} ({\bf q} ,i\Omega)\,,
\label{selfenergy}
\end{equation}
where $i,j$ are sublattice indices. The self-energy for a $\da$-spin electron would be similar to 
that in equation (\ref{selfenergy}) with the 
${\cal G}^{(0)ij}$-spin reversed and $\chi_{-+}$ repaced with $\chi_{+-}$.
The renormalized staggered magnetization at $T=0$ is given by
\begin{equation}
\bar m =-
\frac 1 {N}\sum_{\bm k \sigma}\int_{-\infty}^0\frac{d \om}{2\pi}
\sigma  [Im\,G^{aa}_{\sigma ,{\rm Ret}}(\bm k, \om)-
Im\,G^{bb}_{\sigma ,{\rm Ret}}(\bm k,\om)]\,,
\end{equation}
where $Im\,G^{ij}_{\sigma ,{\rm Ret}}(\bm k, \om)$ stands for the imaginay part
of the retarded Green's function for a spin $\sigma$ electron.

The RPA susceptibility has poles corresponding to the spin waves 
calculated in section 
\ref{collective}, with energy $\approx \vert \phi({\bf k})\vert^2/U$, 
but it also has poles describing a particle-hole continuum of excitations
at higher energies (of order $U$).
In what follows we ignore this particle-hole continuum and 
take into account only 
the contribution from the spin wave poles  to the selfenergy. 
Physically, this means 
that  we shall calculate the magnetization  renormalized by the spin waves. 
To this end, we start by replacing the susceptibility in 
equation (\ref{selfenergy}) by the 
expression 
\begin{equation}
\chi^{(RPA)ij} ({\bf q} , i\omega)= \frac{R^{ij}[\omega({\bf q})]}
{i\omega - \omega({\bf q})} + 
\frac{R^{ij}[-\omega({\bf q})]}{i\omega + \omega({\bf q})}\,,
\label{aldrabice}
\end{equation}
where $R^{ij}[\pm\omega({\bf q})]$ denotes the 
residue of $\chi_{-+}^{(RPA)ij}$ at the spin wave
pole with dispersion $\omega({\bf q})$. Equation (\ref{aldrabice}) 
describes an effective
spin wave propagator.
 After performing the Matsubara frequency summation in 
equation (\ref{selfenergy}) 
we obtain:
\begin{eqnarray}
 \Sigma_{\ua}^{ij}({\bf p} , i\omega) = \frac{U^2}{N}\sum_{\bf q}
&\Big[&\
 \frac{{\rm num}\{ {\cal G}_{\da,-}^{(0)ij}({\bf p - q})\} 
R^{ij}[-\omega({\bf q})]}
{i\omega+\omega({\bf q}) + E_+({\bf p - q})}
-  \frac{{\rm num}\{ {\cal G}_{\da,+}^{(0)ij}({\bf p - q})\} 
R^{ij}[\omega({\bf q})]}
{i\omega-\omega({\bf q}) - E_+({\bf p - q})}\Big]
\label{selfaprox}
\end{eqnarray}
where we have introduced the 
notation ${\rm num}\{ {\cal G}_{\sigma,b}^{(0)ij}\}$ for
the numerators of the  Green's functions, as expressed 
in equations (\ref{gaa}) through  (\ref{coerenc}). 
\begin{figure}[ht]
\begin{center}
\epsfxsize=8cm
\epsfbox{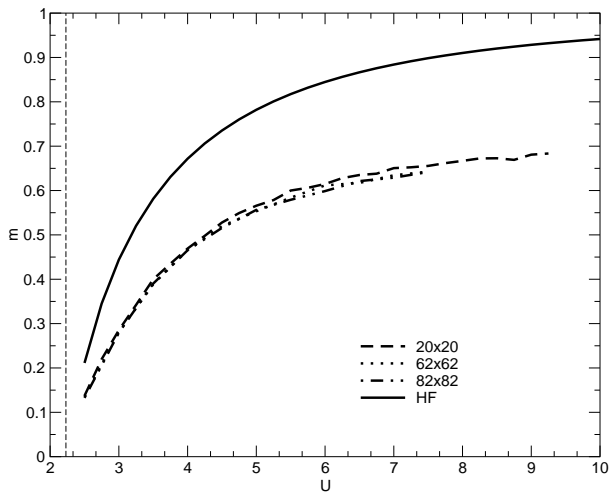}
\end{center}
\caption{The magnetization in the half-filled honeycomb AF layer. 
The continuous
line represents the Hartree-Fock result. Renormalized magnetizations are 
shown for different lattice sizes: 20$\times$20; 62$\times$62; 82$\times$82. 
The vertical dashed line represents
the mean field critical $U$ value at which the magnetic instability develops.}
\label{magnetizations}
\end{figure} 

Figure \ref{magnetizations} we show the 
renormalized magnetization versus $U$. 
The Hartree-Fock magnetization
is also shown in the Figure \ref{magnetizations} for comparison. 
The calculation was performed
for three different lattice sizes. It can be seen that 
convergence does not require a very large number of ${\bf k}$ 
points in the Brillouin Zone. 
This is not surprising because  the Hartree-Fock magnetization itself already
converges to the correct value in  a 40$\times$40 lattice. 
We have also checked that the RPA propagators return the original 
electron density
$n=1$, meaning that no spectral weight was lost in the used approximation for the self
energy.
In the large $U$ limit, 
the renormalized magnetization
saturates at about $67\%$ of  the  (fully polarized) mean field value. 
This is in qualitative 
agreement with the Holstein-Primakoff
result for the  $S=1/2$ Heisenberg model in the honeycomb lattice, 
which predicts a ground state magnetization  of $48\%$.
We should remark, however, that the spin wave spectrum calculated 
within RPA theory has  shown 
much better agreement with experimental results for Mott-Hubbard 
antiferromagnetic insulators 
than the Holstein-Primakoff theory
\cite{la2cuo4,poznan}.

%\begin{table}
%\begin{center}
%\vspace{3mm}
%\begin{tabular}{ccccc}
%\hline
%U& $N=20\times 20$ & $N=62\times 62$ &  $N=82\times 82$ & HF    \\
%\hline
%  2.5 & 0.1371 & 0.13501 &  0.1325 &  0.2112    \\
%  3.0  & 0.2851 & 0.2807    &  0.2794 &   0.444 \\
%  3.5 & 0.3998 &  0.3886 & 0.3902 &  0.5813 \\
%  4.0  & 0.4692 & 0.4659   & 0.4654 &  0.6717  \\
%  4.5 &   0.5278 & 0.5199   & 0.5157  &   0.7353  \\   
%  5.0  & 0.5657  & 0.5532    & 0.5563 & 0.7819  \\
%  5.5 &  0.5999 & 0.5840    &  0.5791 & 0.8173  \\
%6.0  & 0.6147  &  0.6117     & 0.5987  & 0.8448  \\
%6.5 &  0.6351 & 0.6171     & 0.6212 & 0.8665   \\ 
%  7.0 &    0.6504  & 0.6344    & 0.6290 & 0.8840  \\
%7.5 &  0.6555 & 0.6426     & 0.6417 &  0.8983  \\
%8.0  &  0.6669 & 0.6514  &  0.6469       & 0.9102 \\
%\hline
%\end{tabular}
%\end{center} 
%\caption{Renormalized magnetization, as function of $U$, 
%for three different lattice sizes. The column HF  is the Hartree-Fock result.
%\cite{note2} }
%\label{tabela}
%\end{table}

\begin{figure}[ht]
\begin{center}
\epsfxsize=12cm
\epsfbox{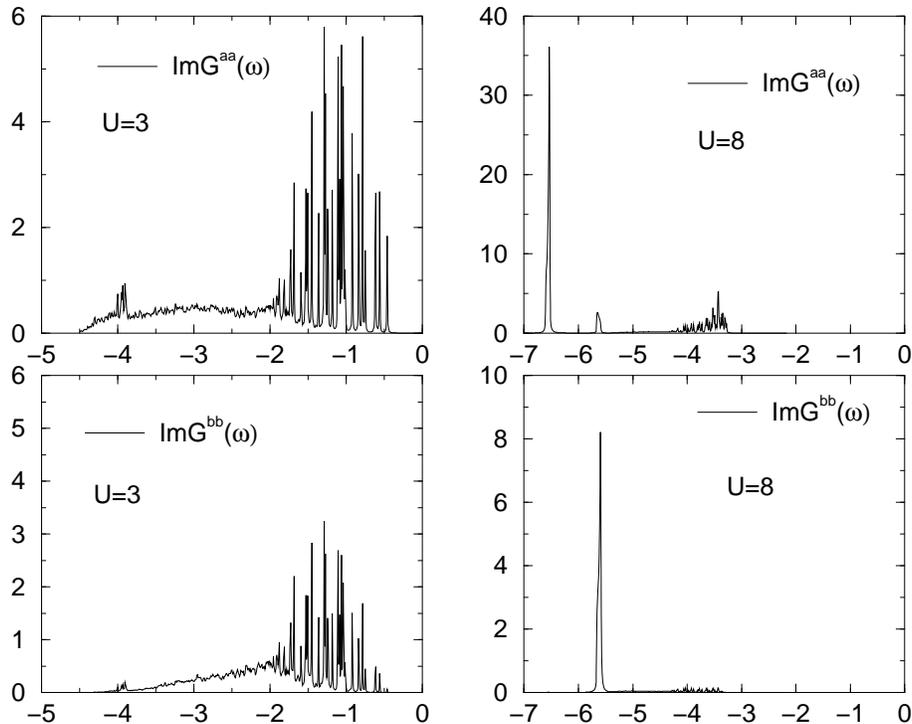}
\end{center}
\caption{Imaginary part of the retarded 
electron Green's function  multipied by $-1$, $-Im G_\ua^{aa(bb)}(\omega)$, versus negative
frequency. The  Green's function
 includes the quantum fluctuations.}  
\label{ImG}
\end{figure}

 In Figure \ref{ImG} we show the imaginary part of the electron's 
Green's function
 at negative frequencies, on both sublattices, for two different 
values of $U$. 
 It is clear that, for strong couplings,  
part of the Hatree-Fock spectral weight is shifted to the bottom of the (negative) energy band.
This shifting of the spectral  weight is responsible for the
renormalization of the staggered magnetization. It is interesting to see that for low
$U$ the spectral weight is most significant at high energy, in the interval [-2,0[, with
a much smaller weight in the interval ]-4,-2[. At a stronger  Hubbard interaction
  most of  the high energy spectral weight (previously in the  interval [-2,0[)
has been displaced  to lower energies and  become localized around
well defined energies, whereas  the 
spectral weight  at intermediate energy  (in the interval ]-4,-2[)  remains essentially
unchanged. Therefore, increasing Hubbard coupling has the efect of displacing the distribution
of  spectral weight from the top to the bottom of the energy  band.

Finally, a comment regarding approximation (\ref{aldrabice}). The commutation
relation between the spin raising and lowering operators,
$$
\sum_{{\bf p},{\bf p'}} [ \  {\hat a}_{{\bf p},\da}^\dagger {\hat a}_{{\bf p}+{\bf q},\ua},
 {\hat a}_{{\bf p'}+{\bf q},\ua}^\dagger  {\hat a}_{{\bf p'},\da}] 
 = \sum_{\bf p} \Big({\hat a}_{{\bf p},\da}^\dagger {\hat a}_{{\bf p},\da} - 
 {\hat a}_{{\bf p},\ua}^\dagger {\hat a}_{{\bf p},\ua}\Big) \,,
$$
is equivalent to the following  relation between the Hartree-Fock  magnetization, $m$, and 
the transverse susceptibilities:
\begin{eqnarray}
\chi_{-+}^{aa} ({\bf q},\tau=0^+)-\chi_{-+}^{aa} ({\bf q},\tau=0^-)  &=&
 \oint_{-i\infty}^{+i\infty} \frac{-idz}{2\pi} \chi_{-+}^{aa}(z)e^{-z 0^+} - 
\oint_{-i\infty}^{+i\infty} \frac{-idz}{2\pi} \chi_{-+}^{aa}(z)e^{z 0^+}
\nonumber  \\  &=&
 -m \,, \label{relation}
\end{eqnarray}
at $T=0$. The integration of the term $e^{-z 0^+}$ ($e^{z 0^+}$) is performed 
along the semi-circular contour on the right (left) half of complex plane. Approximation 
(\ref{aldrabice}) would predict 
\begin{equation}
 R^{bb}[\omega(\vec q)]
-  R^{aa}[\omega(\vec q)] = m\,.
\label{somaregra}
\end{equation}
Indeed, we have checked that our numerical calculation of 
the residues satisfies (\ref{somaregra})
to an accuracy of $1.3\%$.

%%%%%%%%%%%%%%%%%%%%%%%%%%%%
%   Final Remarks          %
%%%%%%%%%%%%%%%%%%%%%%%%%%%%%%%%%%%%%%%%%%%%%%%%%%%%%%%%%%%%%%%%%%%%%%%%%%%%%%%
\section{Final Remarks}
\label{remarks}

In this paper we have studied the magnetic properties of the Hubbard model
in honeycomb layers. Our study focused on the instabilities of the
paramagnetic phase, on the magnetic phase diagram and on the collective
excitations of the half filled phase. Of particular interest is the
fact that it is not possible to describe a true spiraling state in the honeycomb
lattice, as opposed to the usual cubic case. As a consequence, the magnetic spiral
order follows a kind of one dimensional path over the 2D lattice. This kind of ordering,
 here studied at mean field level, may have important consequences to the
study of spin charge separation in 2D lattices. Also interesting,
was the identification of two types of ferromagnetic order, which have eluded
previous studies. For moderate values of $U$ and electron densities not far from the
half filled case, a region of weak ferromagnetism was found to have lower energy
than the more usual Nagaoka ferromagnetic phase.  The renormalization effect of the spin wave
excitations 
on the Hartree Fock magnetization was also studied.
 However, our calculation does not take into 
account  the renormalization of the mean field critical $U$. It is well
known that quantum fluctuations should  induce an increase the value of  $U_c$.
Our calculation cannot capture this effect, since it only takes into account
the effect of well defined spin waves. We believe, however, 
that the calculation can be extended
to include the effect of high-energy damped particle-hole processes leading to
a renormalization of  $U_c$, but this would require a modification of our
numerical calculations and a significant increase of the computational time.
 
%%%%%%%%%%%%%%%%%%%%
%  Acknowledgments %
%%%%%%%%%%%%%%%%%%%%%%%%%%%%%%%%%%%%%%%%%%%%%%%%%%%%%%%%%%%%%%%%%%%%%%%%%%%%%%%
%\section{Acknowledgments}
%One of the authors (D.B.) wishes to thank the coffee-making kit he received as %a 
%present at his birthday.

%%%%%%%%%%%%%%%%%%%%
%    Appendix      %
%%%%%%%%%%%%%%%%%%%%%%%%%%%%%%%%%%%%%%%%%%%%%%%%%%%%%%%%%%%%%%%%%%%%%%%%%%%%%%%
\appendix
%%%%%%%%%%%%%%%%%%%%
%    Appendix A    %
%%%%%%%%%%%%%%%%%%%%%%%%%%%%%%%%%%%%%%%%%%%%%%%%%%%%%%%%%%%%%%%%%%%%%%%%%%%%%%%
\section{Useful expressions for the $U_c$ critical lines at $\bm q=0$}
\label{appendsusuc}
In this appendix, we derive the equations for the critical 
lines from the  static susceptibilities 
($\bm q=\bm 0$ and $\omega=0$). Our starting point is
 the zero order spin-spin susceptibility in  equation  (\ref{chi0}).
The Green's functions  in the paramagnetic region are obtained from 
 equations  (\ref{gaa})-(\ref{gbb}) after setting the magnetization to zero.
 Performing the Matsubara summations in (\ref{chi0}),  the analytical continuation  
and taking the zero frequency limit, we obtain
\begin{eqnarray}
\chi^{(0)aa}_{+-,0}(\bm q,0)&=&\frac 1 4 \sum_{\bm k} ( M_{++}(\bm k,\bm q)+M_{+-}(\bm k,\bm q)+
M_{-+}(\bm k,\bm q)+M_{--}(\bm k,\bm q) ) \\
\chi^{(0)ab}_{+-,0}(\bm q,0)&=&\frac 1 4 \sum_{\bm k} e^{i(\psi_{\bm k-\bm q}-\psi_{\bm k})} 
( M_{++}(\bm k,\bm q)-M_{+-}(\bm k,\bm q)-M_{-+}(\bm k,\bm q)+M_{--}(\bm k,\bm q) ) \\
 M_{\alpha,\beta}(\bm k,\bm q) 
&=&\frac {\theta(E_\alpha (\bm k))-\theta(E_\beta (\bm k -\bm q))} 
{E_\alpha (\bm k)-E_\beta (\bm k -\bm q)}\,,
\end{eqnarray}
where  $\psi_{\bm k}=\arg (\phi_{\bm k})$.
The critical interaction strength, $U_c$ , is given by $U_c/ N = 
[\chi^{(0)aa}_{+-,0} \pm \vert \chi^{(0)ab}_{+-,0} \vert\ ]^{-1}$, 
in the limit $\bm q \rightarrow \bm 0$. 
Expanding all  $\bm q$ dependent quantities around the point $\bm q = \bm 0$ up to first order, 
we obtain
\begin{eqnarray}
\chi^{(0)aa}_{+-,0}(\bm q,0)=\frac 1 4 \sum_{\bm k} \delta(E_{+}(\bm k))+\delta(E_{-}(\bm k))
+\frac {\theta( \vert \phi_{\bm k} \vert - \vert D(\bm k) \vert)} 
{\vert \phi_{\bm k} \vert}+\bm q \cdot ( ... )+... \\
\chi^{(0)ab}_{+-,0}(\bm q,0)=\frac 1 4 \sum_{\bm k} \delta(E_{+}(\bm k))+\delta(E_{-}(\bm k))
-\frac {\theta( \vert \phi_{\bm k} \vert - \vert D(\bm k) \vert)} 
{\vert \phi_{\bm k} \vert}+\bm q \cdot ( ... )+... &.
\end{eqnarray}

Inserting this result in the expression for $U_c$ gives (for $\bm q=0$):
\begin{eqnarray}
\left(\frac {U_c} N\right)^{-1}_+&=&\frac 1 2 \sum_{\bm k} 
\{\delta[E_{+}(\bm k)]+\delta[E_{-}(\bm k)]\} \label{plus}\\
\left(\frac {U_c} N\right)^{-1}_-&=&\frac 1 2 \sum_{\bm k} \frac {\theta( \vert \phi_{\bm k} 
\vert - \vert D(\bm k) \vert)} {\vert \phi_{\bm k} \vert}\label{minus}.
\end{eqnarray}
We  recognize    the density of states, $\rho(\epsilon)=\frac 1 N \sum_{\bm k} 
\{ \delta(E_{+}(\bm k)+\mu-\epsilon)+\delta(E_{-}(\bm k)+\mu-\epsilon) \}$,
 appearing in  equation (\ref{plus}), which is just the Stoner criterion. 
The critical interaction strengths are given by
\begin{eqnarray}
U_{c,+}&=&\frac 2 {\rho(\mu)} \\
U_{c,-}&=&\frac 2 {\frac 1 N \sum_{\bm k} \frac {\theta( \vert \phi_{\bm k} 
\vert - \vert D(\bm k) \vert)} {\vert \phi_{\bm k} \vert}}\,.
\end{eqnarray} 
Note that all $t'$ and $t''$ 
dependence is contained in $D(\bm k)$.
Of course, these equations could also have been obtained by taking the limit
 $m_F, m_{AF} \rightarrow 0$ in  equation (\ref{selfconsmagn}).

%%%%%%%%%%%%%%%%%%%%
%    Appendix B    %
%%%%%%%%%%%%%%%%%%%%%%%%%%%%%%%%%%%%%%%%%%%%%%%%%%%%%%%%%%%%%%%%%%%%%%%%%%%%%%%
\section{Large $U$ results for the susceptibilities and spin waves}
\label{appendsuslarge}

We give asymptotic expressions for the susceptibilities $\chi_{+-}^0(z,{\bf q})$ and 
spin wave dispersion for a half-filled
honeycomb antiferromangetic layer with nearest neighbor hopping. In this case, 
the chemical potential $\mu=0$
and the two 
energy bands are given by $E({\bf k})_{\pm} 
= \pm \sqrt{\Big(\frac {Um} 2\Big)^2 +\vert \phi_{\bm k}\vert^2}$.

The expressions for  coherence factors appearing in the single electron propagators, 
expanded up to second order in $t/U$, are:
\begin{eqnarray}
|A_{\ua,+}({\bf k})|^2 &=& |A_{\da,-}({\bf k})|^2 =|B_{\da,+}({\bf k})|^2 
=|B_{\ua,-}({\bf k})|^2
 \approx \frac{|\phi({\bf k})|^2}{U^2m^2}\\
|A_{\ua,-}({\bf k})|^2 &=& |A_{\da,+}({\bf k})|^2 =|B_{\da,-}({\bf k})|^2
=|B_{\ua,+}({\bf k})|^2
\approx 1-\frac{|\phi({\bf k})|^2}{U^2m^2}\\ 
A_{\da,-}^*({\bf k})B_{\da,-}({\bf k})&=&-A_{\ua,+}^*({\bf k})B_{\ua,+}({\bf k})
\nonumber \\ &=&
A_{\ua,-}^*({\bf k})B_{\ua,-}({\bf k})=-A_{\da,+}^*({\bf k})B_{\da,+}({\bf k})
\approx \frac{\phi^*({\bf k})}{Um}
\end{eqnarray}
 We therefore may use  the aproximate expressions for the 
$\chi_{+-}^0$ susceptibilities:
\begin{eqnarray}
\chi^{(0)aa}(z,{\bf q})&\approx& -\frac{1}{N} \sum_{\bf k}
\frac{1}{z- E({\bf k}) - E({\bf k}+{\bf q})}\Big(1-
\frac{|\phi({\bf k})|^2+|\phi({\bf k}+{\bf q})|^2}{U^2 m^2}\Big)\label{chaa}\\
\chi^{(0)bb}(z,{\bf q})&\approx& \frac{1}{N} \sum_{\bf k}
\frac{1}{z+ E({\bf k}) + E({\bf k}+{\bf q})}\Big(1-
\frac{|\phi({\bf k})|^2+|\phi({\bf k}+{\bf q})|^2}{U^2 m^2}\Big)\\
\chi^{(0)ba}(z,{\bf q})&\approx& \frac{1}{N} \sum_{\bf k}\Big(
\frac{1}{z- E({\bf k}) - E({\bf k}+{\bf q})}-\frac{1}{z+ E({\bf k}) + E({\bf k}+{\bf q})}\Big)
\frac{\phi({\bf k})\ \phi^*({\bf k}+{\bf q})}{U^2 m^2}\\
\chi^{(0)ab}(z,{\bf q})&\approx& \frac{1}{N} \sum_{\bf k}\Big(
\frac{1}{z- E({\bf k}) - E({\bf k}+{\bf q})}-\frac{1}{z+ E({\bf k}) + E({\bf k}+{\bf q})}\Big)
\frac{\phi^*({\bf k})\ \phi({\bf k}+{\bf q})}{U^2 m^2}\label{chab}\,.
\end{eqnarray}
We anticipate that the spin wave energies are of order $z\approx t^2/U$ 
so that we  may use the expansion
$$
\frac{1}{z+ E({\bf k}) + E({\bf k}+{\bf q})}   \approx
\frac{1}{Um} \Big[1-\frac{z}{Um}-\frac{|\phi({\bf k})|^2
+|\phi({\bf k}+{\bf q})|^2}{U^2m^2} + ...\Big]
$$
in equations (\ref{chaa})-(\ref{chab}).
The condition (\ref{det}) for the spin wave dispersion  now takes the form:
\begin{equation}
\frac{z^2}{U^2m^4} = \Big[ 1-\frac{1}{m} +\frac{4}{U^2m^3N}\Big(\sum_{\bf p}
 |\phi({\bf p})|^2\Big) \Big]^2
-\frac{4}{U^2m^6} \Big|\frac{1}{N}
\sum_{\bf p}\phi^*({\bf p})\phi({\bf p}+{\bf q})\Big|^2\,. 
\label{abc}
\end{equation}
But we must take into account that the self-consistent equation for the Hatree-Fock
magnetization, expanded to second order in $t/U$,  is
\begin{equation}
1-\frac{1}{m}  \approx - \frac{2}{U^2m^3N}\Big(\sum_{\bf p}|\phi({\bf p})|^2\Big)
\label{1menos1m}
\end{equation}
Introducing (\ref{1menos1m}) in (\ref{abc}) we finally obtain the spin wave dispersion:
\begin{equation}
z=\omega({\bf q})\approx \frac{2}{Um} \sqrt{\Big(\frac{1}{N}
\sum_{\bf p}|\phi({\bf p})|^2\Big)^2 -
 \Big|\frac{1}{N}
\sum_{\bf p}\phi^*({\bf p})\phi({\bf p}+{\bf q})\Big|^2}\,,
\label{omegaprox}
\end{equation}
which agrees with the result predicted by the Holstein-Primakoff theory.

%%%%%%%%%%%%%%%%%%%%
%    Appendix C    %
%%%%%%%%%%%%%%%%%%%%%%%%%%%%%%%%%%%%%%%%%%%%%%%%%%%%%%%%%%%%%%%%%%%%%%%%%%%%%%%
\section{Holstein-Primakoff analysis of the Heisenberg model}
\label{appendHP}

The Heisenberg Hamiltonian in the honeycomb lattice is given by
\begin{equation}
H=\frac J 2 \sum_{i\in A,\bm \delta}[S^z_iS^z_{i+\bm\delta}+\frac 12 
(S^+_iS^-_{i+\bm \delta}+
S^-_iS^+_{i+\bm \delta})]+
\frac J 2 \sum_{i\in B,\bm \delta}[\tilde S^z_i\tilde S^z_{i+\bm\delta}
+\frac 12 (\tilde S^+_i\tilde S^-_{i+\bm \delta}+
\tilde S^-_i\tilde S^+_{i+\bm \delta})]\,.
\end{equation}
We introduce two sets of operators
\begin{equation}
S^z_i=-a^\dag_i a_i+S\,,\hspace{0.5cm}S^+_i=\sqrt{2S-a^\dag_i a_i}\,a_i\,,
\hspace{0.5cm}S^+_i=a_i^\dag\sqrt{2S-a^\dag_i a_i}\,,
\end{equation}
and
\begin{equation}
\tilde S^z_i=-b^\dag_i b_i+S\,,\hspace{0.5cm}
\tilde S^+_i=\sqrt{2S-b^\dag_i b_i}\,b_i\,,
\hspace{0.5cm}\tilde S^+_i=b_i^\dag\sqrt{2S-b^\dag_i b_i}\,.
\end{equation}
Making the usual linear expansion and introducing the momentum representation
for the bosonic operators, the Hamiltonian can be written as
\begin{equation}
H=-JN_AzS^2+JzS\sum_{\bm k}(a^\dag_{\bm k}a_{\bm k}+b^\dag_{\bm k}b_{\bm k})
+JS\sum_{\bm k}(\phi(\bm k)a_{\bm k}b_{-\bm k}+\phi^\ast(\bm k)
b^\dag_{-\bm k}a^\dag_{\bm k})\,.
\end{equation}
Next we introduce a set of quasiparticle operators defined by
\begin{equation}
a^\dag_{\bm k}=u_{\bm k}\gamma^\dag_{1,\bm k}-v_{\bm k}^\ast\gamma_{2,\bm k}\,,
\hspace{1cm}
b^\dag_{-\bm k}=u_{\bm k}\gamma^\dag_{2,\bm k}-v_{\bm k}^\ast
\gamma_{1,\bm k}\,,
\end{equation}
where the coherence factors obey 
$\vert u_{\bm k}\vert ^2- \vert v_{\bm k}\vert ^2=1$. After introducing the
above transformations in the Hamiltonian we find
\begin{eqnarray}
H&=&-JN_AzS^2+\sum_{\bm k} (2JzS \vert v_{\bm k}\vert ^2 -
JS\phi(\bm k)  v_{\bm k}u_{\bm k}^\ast-
JS\phi^\ast(\bm k)  v_{\bm k}^\ast u_{\bm k})\nonumber\\
&+&  \sum_{\bm k;i=1,2}[JzS(\vert u_{\bm k}\vert ^2+ \vert v_{\bm k}\vert ^2)
- JS\phi(\bm k)  v_{\bm k}u_{\bm k}^\ast-
JS\phi^\ast(\bm k)  v_{\bm k}^\ast u_{\bm k})]\gamma^\dag_{i,\bm k}
\gamma_{i,\bm k}\nonumber\\
&+&  \sum_{\bm k}[(-2JzS v_{\bm k} u_{\bm k}+JS\phi(\bm k)v_{\bm k}
 v_{\bm k} + JS\phi^\ast(\bm k)u_{\bm k}
 u_{\bm k})\gamma^\dag_{1,\bm k}
\gamma^\dag_{2,\bm k}+H.c.]\,,
\end{eqnarray}
which implies the conditions
\begin{eqnarray}
JzS(\vert u_{\bm k}\vert ^2+ \vert v_{\bm k}\vert ^2)
- JS\phi(\bm k)  v_{\bm k}u_{\bm k}^\ast-
JS\phi^\ast(\bm k)  v_{\bm k}^\ast u_{\bm k})&=&\omega(\bm k)\,,\nonumber\\
-2JzS v_{\bm k} u_{\bm k}+JS\phi(\bm k)v_{\bm k}
 v_{\bm k} + JS\phi^\ast(\bm k)u_{\bm k}
 u_{\bm k}&=&0\,.
\end{eqnarray}
The second condition reveals that we can choose $u_{\bm k}$ to be real and
$v_{\bm k}^\ast=\phi(\bm k)\alpha(\bm k)$, with $\alpha(\bm k)$ real.
After some straightforward manipulations we find
\begin{equation}
\omega(\bm k) = JS\sqrt{z^2 -\vert \phi_{\bm k}\vert ^2}\,,
\hspace{1cm} \alpha^2(\bm k)=-\frac 1 {2 \vert \phi_{\bm k}\vert ^2}
+\frac z{2 \vert \phi_{\bm k}\vert ^2}\frac {JS}{\omega(\bm k)}\,.
\end{equation}
The stagered magnetization is given by
\begin{equation}
m=S-\frac 1 {2N_A}\sum_{\bm k}\la a^\dag_{\bm k}a_{\bm k}
+b^\dag_{\bm k}b_{\bm k}\ra=S-\frac 1 {N_A}\sum_{\bm k}
\left(-\frac 1 2 +\frac 1 2 \frac {z } 
{\sqrt{z^2-\vert \phi_{\bm k}\vert ^2}}\right)
-\frac 1 {N_A}\sum_{\bm k}
\frac {z n_B[\omega(\bm k)]} 
{\sqrt{z^2-\vert \phi_{\bm k}\vert ^2}}\,,
\end{equation}
and at zero temperature we assume $n_B[\omega(\bm k)]=0$. Computing the 
integral gives a magnetization value of $0.24$, that is
about 50\% the N\'eel value $\frac 1 2$.

%%%%%%%%%%%%%%%%%%%%
%    bibliography  %
%%%%%%%%%%%%%%%%%%%%%%%%%%%%%%%%%%%%%%%%%%%%%%%%%%%%%%%%%%%%%%%%%%%%%%%%%%%%%%%


\begin{thebibliography}{99}
\bibitem{note2}
The calculations presented in this table were performed at
the GCEP cluster, in the Center of Physics of the University of
Minho.
\end{thebibliography}
\end{document}